# MOLECULAR GEOMETRY OF ALKALOIDS PRESENT IN SEEDS OF MEXICAN PRICKLY POPPY

**Ricardo Gobato** [ricardogobato@seed.pr.gov.br]
*Secretaria de Estado da Educação do Paraná (SEED/PR)*
*Av. Maringá, 290, Jardim Dom Bosco, Londrina/PR, 86060-000, Brasil.*

**Desire Francine Gobato Fedrigo** [desirefg@bol.com.br]
*Panoramic Residence, Rua Uruguai, 1380, ap. 405, Parque Jardim America, Centro, Londrina/PR, 86010-210, Brasil*

**Alekssander Gobato** [alekssandergobato@hotmail.com]
*Faculdade Pitágoras Londrina*
*Rua Edwy Taques de Araújo, 1100, Gleba Palhano, Londrina/PR, 86047-500, Brasil.*

**ABSTRACT**

The work is a study of the geometry of the molecules via molecular mechanics of the main alkaloids found in the seeds of Argemone Mexicana Linn, a prickly poppy, which is considered one of the most important species of plants in traditional Mexican and Indian medicine system. The seeds have toxic properties as well as bactericide, hallucinogenic, fungicide, insecticide, in isoquinolines and sanguinarine alkaloids such as berberine. A computational study of the molecular geometry of the molecules through molecular mechanics of the main alkaloids compounds present in plant seeds is described in a computer simulation. The plant has active ingredients compounds: allocryptopine, berberine, chelerythrine, copsitine, dihydrosanguinarine, protopine and sanguinarine. The studied alkaloids form two groups having similar charge distribution among themselves, which have dipole moments of these two times higher than in the other group.

## Contents

## Introduction

The *Mexican Argemone Linn* popularly known as Mexican poppy, Mexican prickly poppy, thistle or holy thistle [1] is a species of poppy found in Mexico and widespread in various parts of the world. It is an extremely resistant, tolerant to drought and poor soils plant, often being the only vegetation in the soil. It has bright yellow latex, and although toxic to grazing animals, is rarely ingested [2]. The Family *Papaveraceae*, informally known as poppies, is an important ethnopharmacological family of 44 genera and about 760 species of flowering plants. The plant is the source of many types of chemical compounds such as flavonoids, although the alkaloids are the most commonly found. In the pharmaceutical efficacy of certain parts of the plant also show toxic effects [3]. It is used in different parts of the world, for treatment of various diseases, including tumors, warts, skin diseases, rheumatism, inflammations, jaundice, leprosy, microbial infections, malaria [3] agrobacteria [4], mong others, and as a larvicide against the *Aedes aegypti*, the dengue vector [5, 6]. Structurally, polyphenols or phenolic have one or more aromatic rings with hydroxyl groups and may occur as simple and complex molecules. The flavonoids are subgroups of polyphenols [3, 7]. The sanguinarine and berberine have a wide range of biological and / or pharmacological properties, including anti-inflammatory, respiratory stimulation, transient hypotension, convulsions, uterine contraction, antiaritmica property, positive inotropic, adrenocorticotropic hormone activity and analgesic effect [8]. Because of its quaternary nitrogen atom and planar polycyclic structure, sanguinarine and berberine can react with nucleophilic amino acids and anionic groups of different biomolecules, receptors and enzymes. For example, these alkaloids bind to microtubules, inhibit various enzymes including sodium-potassium-ATPase, promoting oxidative phosphorylation and are able to intercalate in DNA regions rich in guanine-cytosine [8]. Contents of alkaloids in the seeds of Argemone mexicana in percentage, total alkaloids 0.13%, diidrosanguinarine 87.00%, sanguinarine 5.00%, berberine 0.57%, protopine 0.34%, chelerythrine 0.12% and coptisine 0.03%. [9]

## 1. The plant

### 1.1 *Argemone Mexicana Linn*

The *Argemone Mexican Linn*, family: *Papaveraceae*, genius: Argemone, species: Mexicana, common names: Amapolas del campo, Bermuda Thistle, Bird-in-the-bush, Brahmadanti, Cardo Santo, Caruancho, Chadron, Flowering Thistle, Gamboge Thistle, Gold Thistle of Peru, Hierba Loca, Jamaican Thistle, Kawinchu, Mexican Prickly Poppy, Mexican Thistle, Mexican Thorn Poppy, Prickly Pepper, Prickly Poppy, Queen Thistle, Satayanasi, Shate, Svarnasiri, Thistle-bush, xate, Yellow Thistle, Zebe Dragon. [10, 11]
Common Indian names:
Hindi: *Shialkanta, Satyanashi*; Gujrati: *Darudi*; Danarese: *Balurakkisa, Datturi, Pirangi, datturi*; Marathi: *Daruri, Firangikote-pavola, dhotara*; Sanskrit: *Brahmadandi, Pitopushpa, Srigalkanta, Svarnakshiri*; Malyalam: *Ponnummattu, Kantankattiri*; Tamil: *Kutiyotti, Ponnummuttai*; Telugu: *Brahmadandicettu*. [12]
An annual with prickly leaves, bright yellow flowers, and bristly capsules containing seeds and resembling black mustard seeds. A native of America, it has run wild into many other countries including India. The seeds yield from 22 to 36 per cent of nauseous, bitter, non-edible oil, which is considered a remedy for skin diseases. In small amounts (1-2 ml.) it is a cathartic and in larger doses it causes violent purging and vomiting. The seeds are sometimes found mixed with black mustard. Adulteration of edible mustard oil with argemone oil is probably responsible for outbreaks of epidermic dropsy [13, 14]. Its presence in concentrations of 0.2 per cent or lessis detected by the rich orange-red colour which appears when concentrated nitric acid is added to the oil or its mixtures, or by the ferric chloride test [15, 16]. The plant contains barberine and protopine [17]. The seeds also are considered to have a medicinal value, as a laxative, emetic, expectorant and demulcent; taken in large quantities, they are said to be poisonous. The yellow juice which exudes when the plant is damaged is used externally in scabies, dropsy, jaundice, cutaneous affections and ophthalmic. The oil is considered to be a purgative and is also used for cutaneous affections. [18]

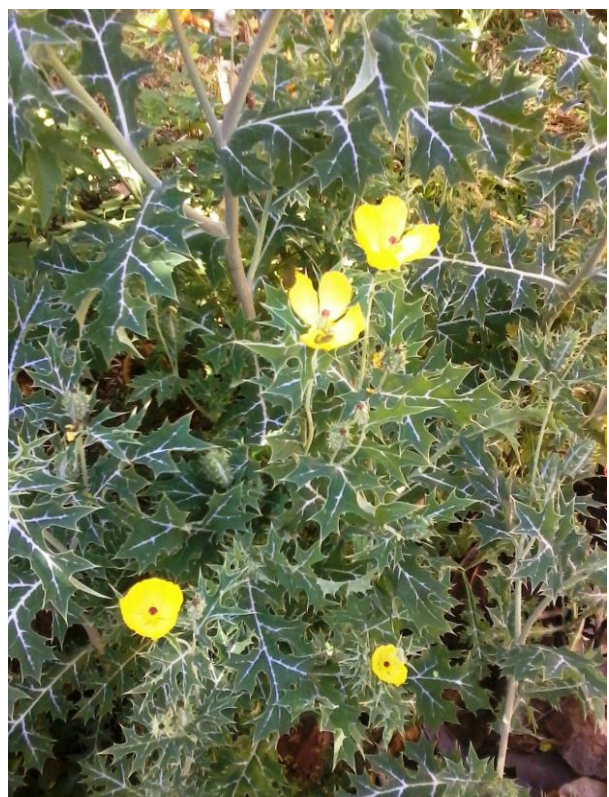

**Figure 1.** Photo *papaveraceae* plant, *Argemone Mexicana Linn*

### 1.2 History

The *Florentine Codex*, in the sixteenth century, indicated for sore eyes. In the same century, *Francisco Hernandez* he says: "evacuates all moods, mainly pituitous and damaging the joints, cure inflammation of the eyes, is effective against fevers accesses, heals ulcers sexual parts, scabies, dissolves clouds eyes, consume superfluous flesh and calm the pain of migraine." In the late eighteenth century, *Vicente Cervantes* tells that "purges pituitous moods, relieves inflammation of the eyes and dissipates the clouds which begin to form in them." [11] The *Mexican Society of Natural History* records it as antidiarrheal, anti-dysentery, anti-gonorrhoeal, astringent, eye diseases, diuretic, emeto-cathartic, hypnotic, chest, hair tonic and analgesic, for cafalalgias, dermatosis. Later, *Francisco Flores* slogan use for simple conjunctivitis, combat the emerging leucome, nephelion eyes, chemosis and as a sedative. In the same years that *Flores, Eleuterio Gonzalez* appointment to cure clouds corneal against headaches as work as a lever and soothing to wash the head and thus is bornhair, and as a drastic purgative[11]. In the twentieth century, *Narciso Souza* describes their use for diseases of the pancreas liver, lack of appetite, as an emetic and purgative, for eye inflammation and skin diseases. Make attributed pectoral properties and soporific. Around the same time that *Souza, Luis Cabrera* registers as antitussive, hypnotic and cough whooping. Finally, the *Pharmaceutical Society of Mexico* mentions its use as anti-escabiotic, antispasmodic, antitussive, cathartic, dermatosis, emeto-cathartic, hypnotic, pectoral and sedative. [11]

### 1.3 Ethnobotany and anthropology

The *Argemone Mexicana Linn* is recommended to treat eye problems such as pain, itching, plant and stains inflammation; although it used primarily cataracts, directly applying the latex fresh or fomentation of decoction of the bark for 5 or 6 days. Milk (latex) applies cool nights on eyelid or inside the eye to remove the sting, and mixed with juice mesquite gets in drops removing eye clouds. To relieve deafness, balls (fruits) are ground, they are placed in a cloth or cotton, and placed in the ear. In *Michoacan* used for *tsandukus* in *Purepecha* is the name given to the disease eye, manifested by excessive secretion, caused by a sudden change in temperature. In Here it may be sufficient one application. Likewise it applied in irritated eyes, although it is suggested put morning. [11]

The root is used, as a poultice, to relieve pain lung, l condition caused by excess work (recognized because his back hurts and feels warm). When it suffers cough caused by cold, not you can breathe and chest pain, then a tea made from the flowers is taken. [11]

It is used to bring out the *chincual*, ie, hives or rashes located in various parts of the body the children. To this end, they were bathing with a decoction of the plant. Others describe the chincual as irritation (redness) of the anus in children. In this case, leaves and stem regrind in a bucket of water to give baths seat. Scabies, can be used fresh and dried grass seed or seed oil. Furthermore, in water where the roots were boiled, often bathe drinkers (alcohol) that have irritated skin. For use as a laxative, the seeds are boiled. It is also reported useful in the treatment of bile, toothache, colic in children (newborn colic V.) expulsion of placenta, flow, wounds and ulcers, kidney pain, diabetes, skin infections, pimples, spots, rashes, inflammation, malaria, convulsions, spasms, infection and bleeding, as a purgative and healing. [10, 11]

### 1.4 Current use

The *Argemone Mexicana Linn* is used as a medicinal plant in several countries. In Mexico the seeds are considered an antidote to the poison of the snakes in India while the smoke of the seeds are used to relieve toothache. The extract fresh seeds contain substances that dissolve the protein and is effective in the treatment of warts, herpes sores, skin infections, skin diseases, itching and also in the treatment of dropsy and jaundice [19]. Also in India the plant is used in *Ayurvedic* and *Unani* medicine for the treatment of a wide range of diseases including malaria. The use against malaria is also widespread in several African countries, including Benin, Mali and Sudan [20]. There is no documentation of the psychoactive properties the plant. From the information contained in the websites it shows that the leaves are smoked as a substitute for *marijuana* or used for making tea, a cigarette accompanying. The *Argemone Mexicana Linn* is sold in the form of resin, mixed with other herbs. It is suggested to mix it with tobacco and smoking it in order to obtain a relaxing effect and flavoring tobacco. For a stronger effect the resin is used in pure bong or vaporized. [10]

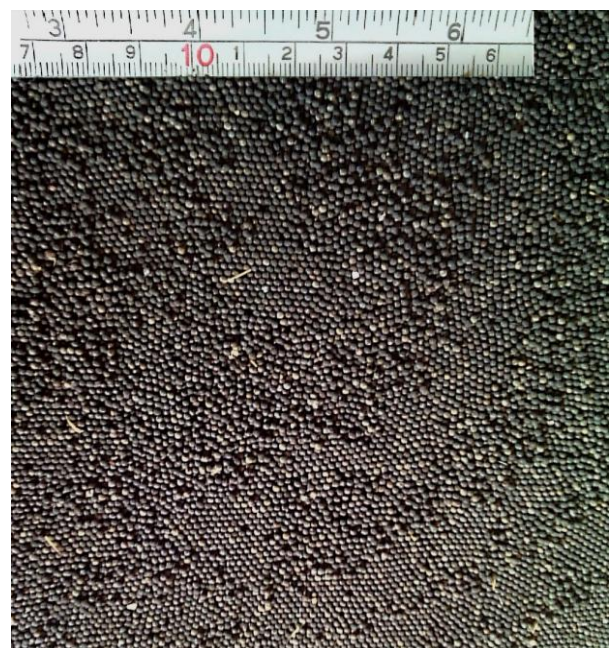

**Figure 2.** Photo of seeds *Argemone Mexicana Linn*.

### 1.5 Toxicity

The seeds of the Mexican reminiscent in appearance and size of those black mustard (*Brassica nigra*). The mustard seed oil is widely used in Indian cuisine and a popular accidental ingestion of this oil adulterated with oil seeds of *Argemone Mexicana Linn* causes dropsy epidemic characterized by abnormal accumulation of fluid in tissues and body cavities[21, 22, 23]. Outbreaks of the disease have been reported in Mauritius, in Fiji, Madagascar and South Africa [21, 22]. Idropisia epidemic studies indicate that an dulteration of oil of Argemone in mustard oil greater than 1% is sufficient to produce clinical symptoms. The toxicity of plant is mainly due to sanguinarine, which seems to be 2.5 times more toxic than the other toxic alkaloid, the dihidrosanguinarina, though both are interconvertible via a simple reaction of oxidation-reduction. The reported toxicity in animal model for sanguinarine, berberine, protopine and cheleryhtrine. [23]

# 2. Fundamentals

## 2.1 Introduction to Molecular Mechanics

The "mechanical" molecular model was developed out of a need to describe molecular structures and properties in as practical a manner as possible. The range of applicability of molecular mechanics includes:

• Molecules containing thousands of atoms.
• Organics, oligonucleotides, peptides, and saccharides (metallo-organics and inorganics in some cases).
• Vacuum, implicit, or explicit solvent environments.
• Ground state only.
• Thermodynamic and kinetic (via molecular dynamics) properties.

The great computational speed of molecular mechanics allows for its use in procedures such as molecular dynamics, conformational energy searching, and docking. All the procedures require large numbers of energy evaluations. Molecular mechanics methods are based on the following principles:

• Nuclei and electrons are lumped into atom-like particles.
• Atom-like particles are spherical (radii obtained from measurements or theory) and have a net charge (obtained from theory).
• Interactions are based on springs and classical potentials.
• Interactions must be preassigned to specific sets of atoms.
• Interactions determine the spatial distribution of atom like particles and their energies. Note how these principles differ from those of quantum mechanics. [24, 25, 26, 27]

In short the goal of molecular mechanics is to predict the detailed structure and physical properties of molecules. Examples of physical properties that can be calculated include enthalpies of formation, entropies, dipole moments, and strain energies. Molecular mechanics calculates the energy of a molecule and then adjusts the energy through changes in bond lengths and angles to obtain the minimum energy structure. [25, 26, 27]

## 2.2 Steric Energy

A molecule can possess different kinds of energy such as bond and thermal energy. Molecular mechanics calculates the steric energy of a molecule – the energy due to the geometry or conformation of a molecule. Energy is minimized in nature, and the conformation of a molecule that is favored is the lowest energy conformation. Knowledge of the conformation of a molecule is important because the structure of a molecule often has a great effect on its reactivity. The effect of structure on reactivity is important for large molecules like proteins. Studies of the conformation of proteins are difficult and therefore interesting, because their size makes many different conformations possible.

Molecular mechanics assumes the steric energy of a molecule to arise from a few, specific interactions within a molecule. These interactions include the stretching or compressing of bonds beyond their equilibrium lengths and angles, torsional effects of twisting about single bonds, the Van der Waals attractions or repulsions of atoms that come close together, and the electrostatic interactions between partial charges in a molecule due to polar bonds. To quantify the contribution of each, these interactions can be modeled by a potential function that gives the energy of the interaction as a function of distance, angle, or charge 1,2. The total steric energy of a molecule can be written as a sum of the energies of the interactions:

$$E_{se} = E_{str} + E_{bend} + E_{str-bend} + E_{oop} + E_{tor} + E_{VdW} + E_{qq} \quad (1)$$

The steric energy, bond stretching, bending, stretch-bend, out of plane, and torsion interactions are called bonded interactions because the atoms involved must be directly bonded or bonded to a common atom. The Van der Waals and electrostatic (qq) interactions are between non-bonded atoms [25, 26, 27].

As mentioned above, the expression for the potential energy of a molecular system that is used most frequently for simple organic molecules and biological macromolecules is the following:

$$V(r) = \sum_{bonds} \frac{k_d}{2}(d - d_o)^2 +$$

$$\sum_{angles} \frac{k_\theta}{2}(\theta - \theta_o)^2 +$$

$$\sum_{dihedrals} \frac{k_\phi}{2}(1 + \cos(\phi - \phi_o)) +$$

$$\sum_{impropers} \frac{k_\psi}{2}(\psi - \psi_o)^2 +$$

$$\sum_{\substack{no-bonded \\ pairs(i,j)}} 4\varepsilon_{ij} \left[ \left(\frac{\sigma^{ij}}{r_{ij}}\right)^{12} + \left(\frac{\sigma^{ij}}{r_{ij}}\right)^{6} \right] +$$

$$\sum_{\substack{no-bonded \\ pairs(i,j)}} \frac{q_i q_j}{\varepsilon_D r_{ij}}$$

(1)

### 2.3 Geometry Optimization

The dynamic was held in Molecular Mechanics Force Field (Mm+), equation (1), computed geometry optimization molecular at algorithm Polak-Ribiere [28], conjugate gradient, at the termination condition: RMS gradient [29] of 0,1 kcal/A.mol or 405 maximum cycles in vacuum. Molecular properties: electrostatic potential 3D mapped isosurface, mapped function range, minimum 0.065 at maximum 0.689, display range legend, from positive color red to negative color blue, total charge density contour value of 0.05, gourand shaded surface.

## 3. Chemical formula and physico-chemical properties of the active ingredients

### 3.1 Allocryptopine
Allocryptopine is a bioactive alkaloid.

CAS No. 485-91-6
Chemical Name: Allocryptopine
Synonyms: Fagarine i; alpha-allocryptopine; beta-Allocryptopine; gamma-homochelidonine; Thalictrimine; allo-Cryptopine; alpha-Fagarine; Fagarine I; alpha-Allocryptopine; beta-Homochelidonine; 5,7,8,15-tetrahydro-3,4-dimethoxy-6-methyl(1,3)benzodioxole(5,6-e)(2)benzazecin-4(6H)-one.
Molecular Formula: $C_{21}H_{23}NO_5$
Molar mass: 369.41 g/mol
Melting point: 160-161℃
Solubility: soluble in alcohol, chloroform, ether, ethyl acetate and dilute acids
UVmax: 232, 284 nm. [30, 31]

### 3.2 Berberine
Berberine is a quaternary ammonium salt from the protoberberine group of isoquinoline alkaloids. It is found in such plants as *Berberis* [e.g. *Berberis aquifolium* (Oregon grape), Berberis vulgaris (barberry), *Berberis aristata* (tree turmeric*)]*, *Hydrastis canadensis* (goldenseal), *Xanthorhiza simplicissima* (yellowroot), *Phellodendron amurense* [32] (Amur cork tree), *Coptis chinensis* (Chinese goldthread or Huang Lian Su), *Tinospora cordifolia, Argemone mexicana* (prickly poppy), and *Eschscholzia californica* (Californian poppy). Berberine is usually found in the roots, rhizomes, stems, and bark [32]. Berberine was supposedly used in China as a broad-spectrum anti-microbial medicine by *Shennong* around 3000 BC. This first recorded use of Berberine is described in the ancient Chinese medical book *The Divine Farmer's Herb-Root Classic*. Due to Berberine's strong yellow color, Berberis species were used to dye wool, leather, and wood. Wool is still dyed with berberine today in northern India. Under ultraviolet light, berberine shows a strong yellow fluorescence [33], so it is used in histology for staining heparin in mast cells [31]. As a natural dye, berberine has a colour index of 75160.

CAS No. 2086-83-1
Chemical Name: Berberine Synonyms: berberin; umbellatine; 5,6-dihydro-9,10-dimethoxybenzo[g]-1,3-benzodioxolo[5,6-a]quinolizinium.
Molecular Formula: $[C_{20}H_{18}NO_4]^+$
Molar mass: 336.36122 g/mol
Appearance: yellow solid
Melting point: 145 °C (293 °F; 418 K)
Solubility in water: slowly soluble
UVmax: 265, 343 nm. [30, 33]

### 3.3 Chelerythrine
Chelerythrine is a benzophenanthridine alkaloid present in the plant Chelidonium majus (greater celandine). It is a potent, selective, and cell-permeable protein kinase C inhibitor[34, 35]. It is also found in the plants *Zanthoxylum clava-herculis* and *Zanthoxylum rhoifolium*, exhibiting antibacterial activity against *Staphylococcus aureus* and other human pathogens. [36, 37]

CAS No. 3895-92-9
Chemical Name: Chelerythrine
Synonyms:1,2-dimethoxy-12-ethyl[1,3]benzodioxolo[5,6-c]phenanthridin-12-ium; 3)benzodioxolo(5,6-c)phenanthridinium,1,2-dimethoxy-12-methyl;5,6-dihydro-9,10-dimethoxybenzo[g]-1,3-benzodioxolo[5,6-a]quinolizinium;Toddalin.
Molecular Formula: $[C_{21}H_{18}NO_4]^+$
Molar mass: 348.37192 g/mol
Melting point: 200-206 °C.
Solubility: soluble in dimethyl sulfoxide (DMSO), ethyl alcohol, petroleum ether. Not soluble in water.
UVmax: 226, 283, 320 nm. [34]

### 3.4 Copsitine

Coptisine is an alkaloid found in Chinese goldthread (Coptis chinensis) [38]. Famous for the bitter taste that it produces, it is used in Chinese herbal medicine along with the related compound berberine for treating digestive disorders caused by bacterial infections. Also found in Greater Celandine and has also been detected in *Opium* [39]. Coptisine has been found to reversibly inhibit Monoamine oxidase A in mice, pointing to a potential role as a natural antidepressant [40]. However, this may also imply a hazard for those taking other medications or with a natural functional disorder in Monoamine oxidase A. Coptisine was found to be toxic to larval brine shrimp and a variety of human cell lines, potentially implying a therapeutic effect on cancer or alternatively a generally toxic character. The same authors illustrate a four-step process to produce Coptisine from Berberine. [41]

CAS No. 3486-66-6
Chemical Name: Copsitine
Synonyms: 6,7-Dihydro-bis(1,3)benzodioxolo (5,6-a:4′,5′-g)quinolizinium; Coptisin; 5,6-Dihydro-2,3:9,10-bis(methylenedioxy)dibenzo[a,g]quinolizinium; 6,7-Dihydrobis[1,3]benzodioxolo[5,6-a:4′,5′-g] quinolizinium; Bis[1,3]benzodioxolo[5,6-a:4′,5′-g]quinolizinium,6,7-dihydrobis[methylenedioxy]protoberberine 7,8,13,13a-Tetradehydro-2,3-9,10-bis (methylenedioxy)berbinium;
6,7-Dihydro[1,3]dioxolo[4,5-G][1,3] dioxolo[7,8]isoquino[3,2-A]isoquinolin-5-Ium.
Molecular Formula: $[C_{19}H_{14}NO_4]^+$
Molar mass: 320.319 g/mol
Melting point: 160-161 °C
Solubility: very slightly soluble in water, partially soluble in alcohol, soluble in alkali.
UVmax: 229, 244, 267, 353 nm. [30, 34, 42]

### 3.5 Dihydrosanguinarine

CAS No. 3606-45-9
Chemical Name: Dihydrosanguinarine
Synonyms:13-Methyl-13,14-dihydro-[1,3]dioxolo[4′,5′:4,5]benzo[1,2-c][1,3]dioxolo[4,5-i]phenanthridine;
13,14-Dihydro-13-methyl-[1,3]benzodioxolo[5,6-c]-1,3-dioxolo[4,5-i]phenanthridine; dihydroavicine; Hydrosanguinarine
Molecular Formula: $C_{20}H_{15}NO_4$
Molar mass: 333.3374 g/mol
Melting point: 188-189 °C.
Solubility: they are not present in the literature data relating to the solubility.
UVmax: 237, 284, 322 nm in ethanol. [30, 34]

### 3.6 Protopine

Protopine is a benzylisoquinoline alkaloid occurring in opium poppy [43], Corydalis tubers [44] and other plants of the family *papaveraceae*, like *Fumaria officinalis* [45]. It has been found to inhibit histamine H1 receptors and platelet aggregation, and acts as an analgesic. [46]

CAS No. 130-86-9
Chemical Name: Protopine
Synonyms: 7-Methyl-6,8,9,16-tetrahydrobis[1,3] benzodioxolo[4,5-c:5′,6′-g]azecin-15(7H)-one; Fumarine; Biflorine; Corydinine; Macleyine; Protopin; 4,6,7,14-tetraidro-5-metil-bis(1,3)
-benzodiossolo(4,5-c-5′,6′-g)azecin-13(5H)-one.
Molecular Formula: $C_{20}H_{19}NO_5$
Molar mass: 353.369 g/mol
Appearance: white crystals
Density: 1.399 g/cm$^3$
Melting point: 208 °C (406 °F; 481 K)
Solubility: soluble in ethyl acetate, carbon bisulfide, benzene petroleum, ether. Practically insoluble in water.
Solubility in chloroform: 1:15
UVmax: 239, 291, in ethyl alcohol 95% = 293 nm. [30, 34, 42]

### 3.7 Sanguinarine

Sanguinarine is a toxic quaternary ammonium salt from the group of benzylisoquinoline alkaloids. It is extracted from some plants, including bloodroot (*Sanguinaria canadensis*), Mexican prickly poppy Argemone Mexicana [47], *Chelidonium majus* and *Macleaya cordata*. It is also found in the root, stem and leaves of the opium poppy but not in the capsule. Sanguinarine is a toxin that kills animal cells through its action on the Na+-K+-ATPase transmembrane protein [48]. Epidemic dropsy is a disease that results from ingesting sanguinarine [9]. If applied to the skin, sanguinarine kills cells and may destroy tissue. In turn, the bleeding wound may produce a massive scab, called an eschar. For this reason, sanguinarine is termed an escharotic. [49]

CAS No. 2447-54-3
Chemical Name: Sanguinarine
Synonyms:13-Methyl-[1,3]benzodioxolo[5,6-c]-1,3-dioxolo[4,5-i]phenanthridinium
Molecular Formula: $C_{20}H_{14}NO_4$
Molar mass: 332.3 g/mol
Melting point: 205-215 °C
Solubility: soluble in alcohol, chloroform, acetone, ethyl acetate.
UVmax: 234, 283, 325 nm in methyl alcohol. [30, 31, 34, 42]

**Table 1.** Total Charge Density of Alkaloids

|  | Allocryptopine | Berberine | Chelerythrine | Copsitine | Dihydrosanguinarine | Protopine | Sanguinarine |
|---|---|---|---|---|---|---|---|
| Blue($\delta^+$) | 0,338 | 0,679 | 0,689 | 0,680 | 0,332 | 0,333 | 0,338 |
| Red($\delta^-$) | 0,066 | 0,191 | 0,143 | 0,143 | 0,065 | 0,065 | 0,065 |
| Δδ | 0,272 | 0,488 | 0,546 | 0,537 | 0,267 | 0,268 | 0,273 |

## 4. Discussion and conclusions

Figure 1 is a plan photo *Argemone Mexicana Linn*, with green leaves and bright yellow flowers and fruits. Figure 2 shows a photo of plant seeds with scale in centimeters and inches. Figures 3 to 9, the molecular distribution of load densities and molecular geometry of the main alkaloids of the plant after a molecular dynamics Mm + using *HyperChem 7.5 Evaluation* software [50]. Figures 3 to 9, the molecules of Allocryptopine, Berberine, Chelerythrine, Copsitine, Dihydrosanguinarine, Protopine and Sanguinarine, respectively. Table 1 shows the density distribution of charges to the main alkaloid of plant studied. The alkaloids Allocryptopine, Dihydrosanguinarine, Protopine and Sanguinarine - group 1 have density similar negative and positive charges, as well as Berberine, Chelerythrine, Copsitine, group 2. Group 1, has a charge density difference twice lower, relation to the group 2. The group 2 when the dipole has to be two times greater than group 1. The main sites of positive charge density of group 2 are oxygens, nitrogens as are the negative sites at the ends, and molecular center, respectively. Already the main local density of positive charges from Group 1 are the hydrogens atoms distributed by molecular contours, and the negative oxygens atoms in its longitudinal ends, and cross for Allocryptopine and Protopine.

The studied alkaloids form two groups having similar charge distribution among themselves, which have dipole moments of these two times higher than in the other group.

## Figures

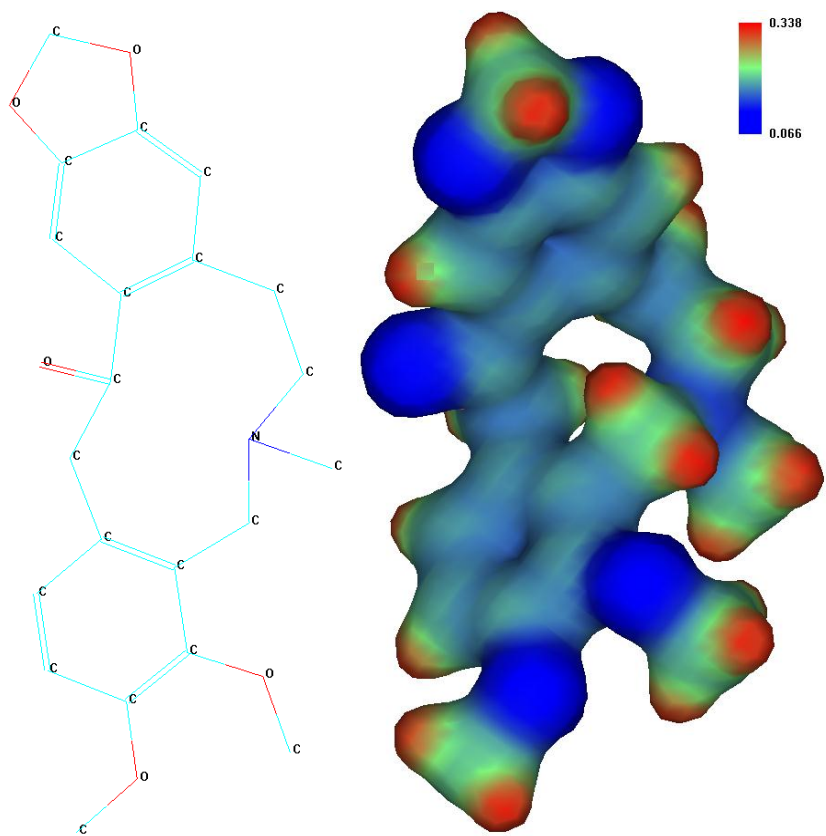

**Figure 3.** Allocryptopine.

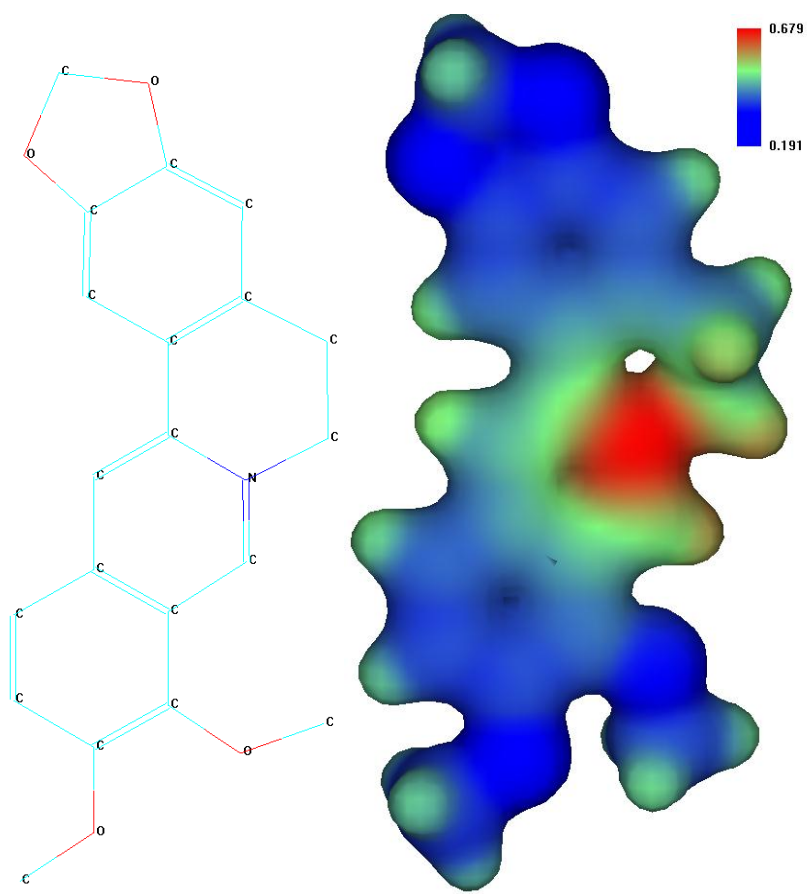

**Figure 4.** Berberine.

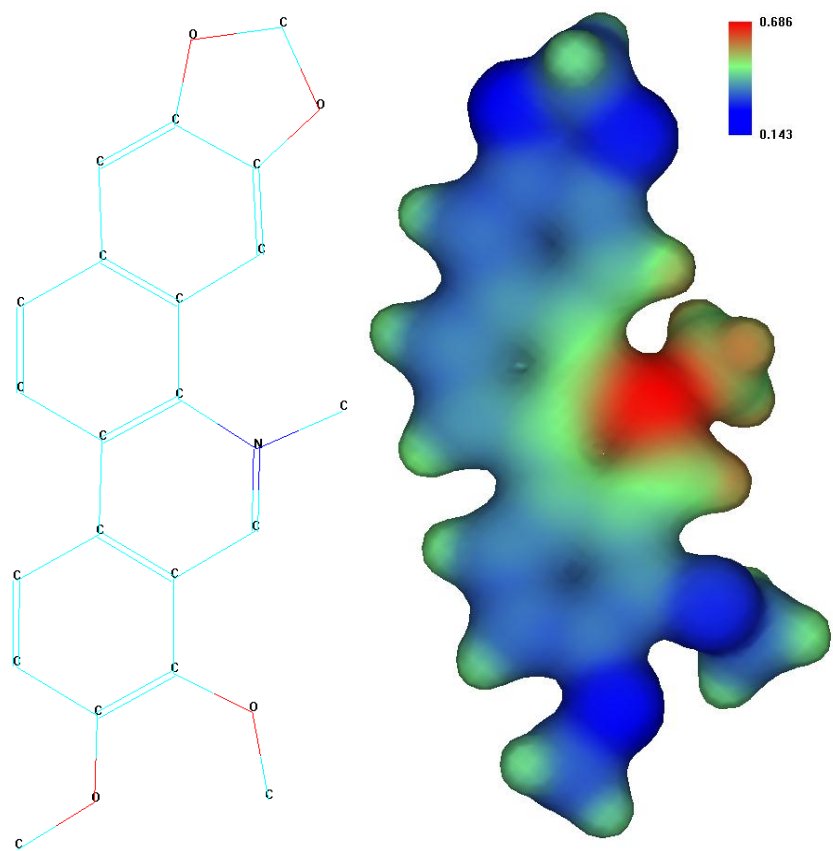

**Figure 5.** Chelerythrine.

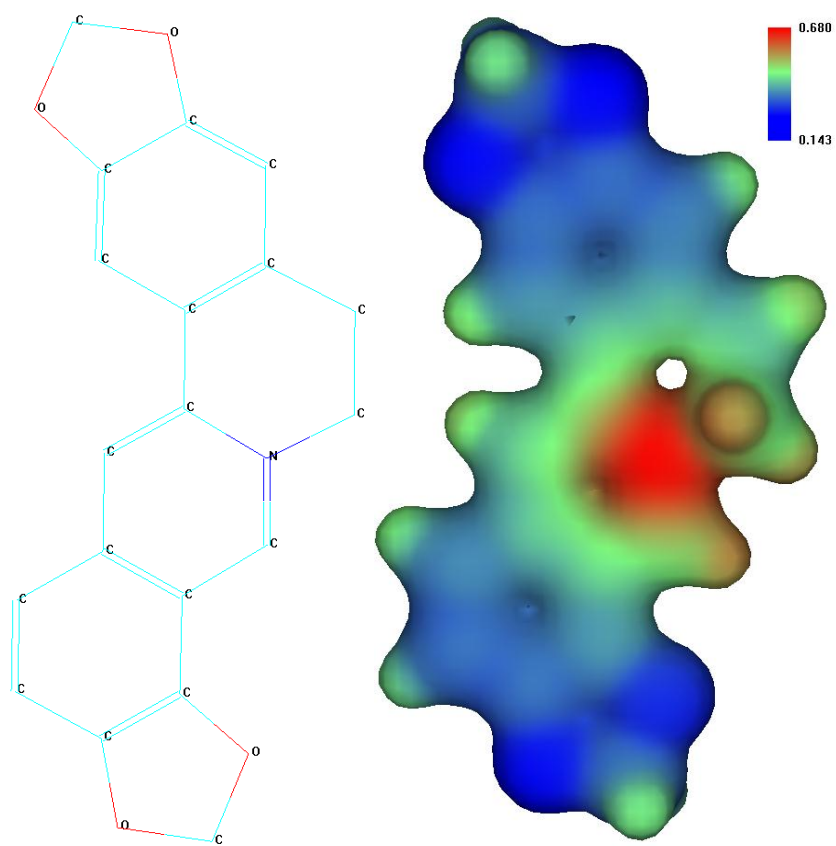

**Figure 6.** Copsitine.

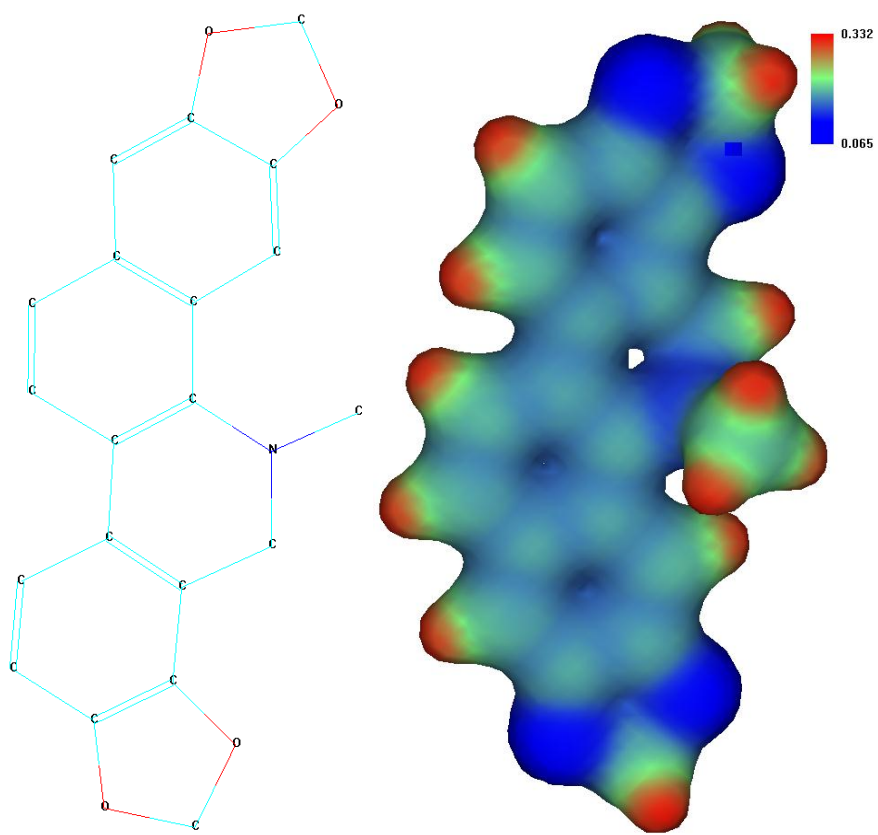

**Figure 7.** Dihydrosanguinarine.

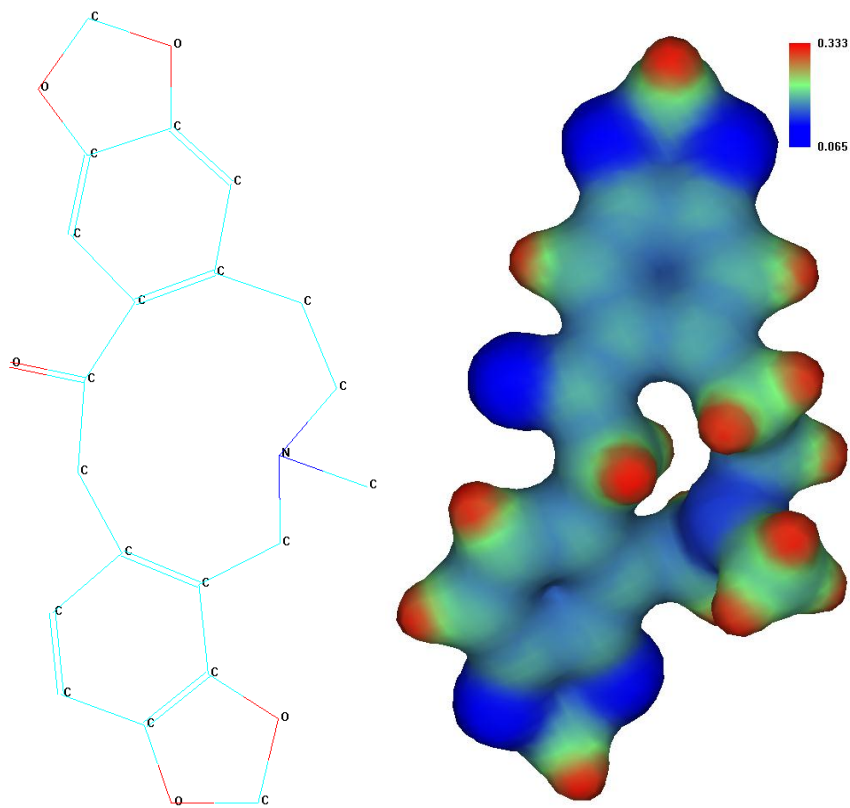

**Figure 8.** Protopine.

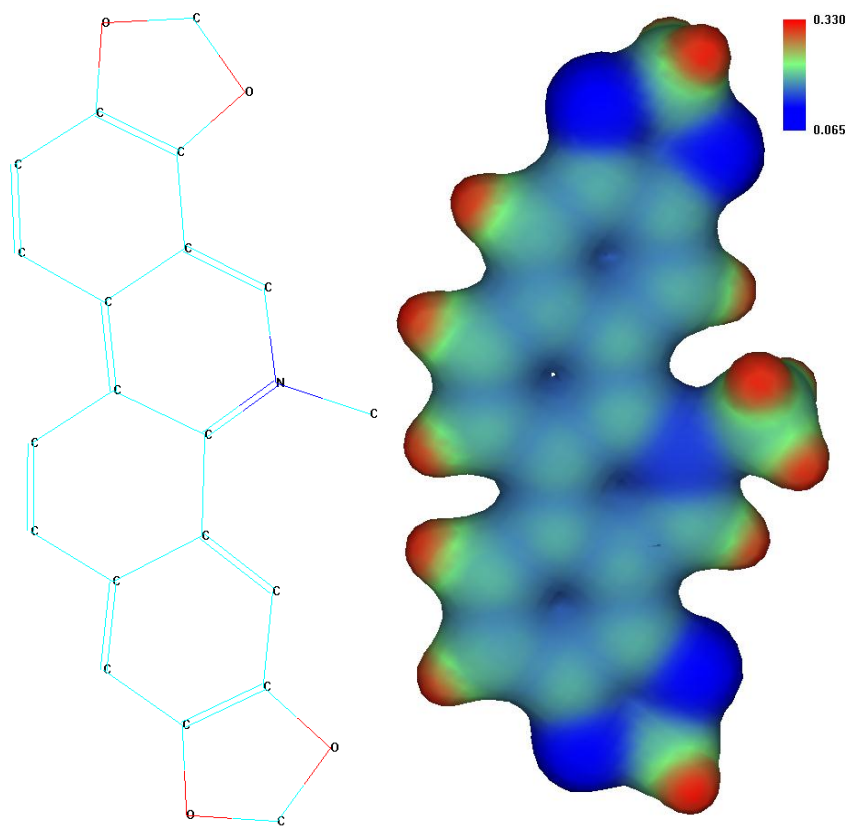

**Figure 9.** Sanguinarine.

## Acknowledgments

Thank GOD